\documentclass[aps,pra,twocolumn,showpacs]{revtex4}
\usepackage{graphicx}
\usepackage{amsmath}
\usepackage{amsfonts}
\usepackage{amssymb}
\newcommand{\ket}[1]{|#1\rangle}
\newcommand{\bra}[1]{\langle #1|}
\newcommand{\Tr}{\text{Tr}}
\newcommand{\im}{\text{Im}}
\newcommand{\re}{\text{Re}}
\begin{document}
\title{Hidden parameters in open-system evolution unveiled by geometric phase}
\author{Patrik Pawlus$^1$ and Erik Sj\"oqvist$^{1,2}$}
\affiliation{$^1$Department of Quantum Chemistry, Uppsala University,
Box 518, Se-751 20 Uppsala, Sweden \\
$^2$Centre for Quantum Technologies, National University
of Singapore, 3 Science Drive 2, 117543 Singapore, Singapore}
\begin{abstract}
We find a class of open-system models in which individual quantum trajectories
may depend on parameters that are undetermined by the full open-system evolution.
This dependence is imprinted in the geometric phase associated with
such trajectories and persists after averaging. Our findings indicate a potential
source of ambiguity in the quantum trajectory approach to open quantum
systems.
\end{abstract}
\maketitle
\section{Introduction}
\label{sec:introduction}
Closed quantum systems evolving deterministically under some Hermitian Hamiltonian
is an idealized description that at best approximates real laboratory
experiments. In fact, all quantum systems undergo open-system
effects induced by entanglement with environmental degrees of freedom;
effects that may be detrimental in various quantum information protocols
in which coherence is an essential ingredient \cite{nielsen00}. This feature
has led to a revived interest in the theory of open quantum systems and
how to deal with open-system effects by different types of error control to
achieve error resilient quantum information processing
\cite{shor95}.

Geometric and holonomic quantum computation, first conceived in Ref.
\cite{zanardi99} and experimentally demonstrated in Ref. \cite{jones00},
is an approach to error control that has attracted considerable attention
recently. In its simplest variant, it makes use of the Abelian geometric
phase \cite{berry84} to construct quantum logical gates acting on one or
two quantum mechanical bits (qubits) \cite{ekert00}. These gates may be
used to build quantum Boolean networks and may be combined with other
error resilient methods to perform robust quantum computation
\cite{wu05,oreshkov09}. The need to understand the error resilience
of geometric and holonomic quantum computation has led to proposals
for the geometric phase of open quantum systems
\cite{carollo03,tong04,marzlin04,sarandy06}.

Here, we examine the idea in Ref.~\cite{carollo03} (see also Ref. \cite{carollo04a})
to associate geometric phases of individual quantum trajectories in
quantum jump unravelings of Lindblad-type open-system evolution
\cite{plenio98}. This approach involves only pure state geometric phases,
which may relate to the geometry of the full open-system evolution by
some averaging over trajectories \cite{sjoqvist06}.

The trajectory-based geometric phase simplifies the analysis of the robustness
of geometric and holonomic quantum computation \cite{cen04,fuentes05,moller08}.
The idea is that for weak influence of the environment, it suffices to consider
the lowest order, no-jump trajectory to evaluate error resilience. Here, we show
that this geometric phase may in
certain cases lead to different predictions regarding the resilience of geometric
and holonomic quantum computation to open-system effects. This result indicates a
potential source of ambiguity in the no-jump approach to analyze weak open-system
effects.

The problem of how to define open-system geometric phases by averaging over quantum
trajectories has been addressed in Refs. \cite{bassi06,buric09}. These works employ
quantum state diffusion (QSD) \cite{gisin92}, which is a form of stochastic unravelings
of the Lindblad evolution consisting of continuous, Brownian-type trajectories in
state space.

In Ref. \cite{bassi06}, the averaged geometric phase associated with the full
nonlinear form of the QSD equation \cite{gisin92} was examined. It was found that
this phase is not invariant under unitary rotations $L_m \rightarrow \sum_n V_{mn}L_n$
of the Lindblad operators $L_m$. On the other hand, Ref. \cite{buric09} demonstrated
that this noninvariance would disappear if the averaged geometric phase is instead
associated with the linearized version of QSD \cite{goetsch94}, provided the system
starts in a pure state. Based on this result, it was claimed in Ref. \cite{buric09}
that the linearized QSD approach provides a uniquely defined geometric phase of open
systems. Here, we demonstrate the existence of a class of
Markovian open-system evolutions for which the linearized QSD geometric phase may
change under symmetry transformations of the full Lindblad evolution.

The outline of the paper is as follows. In the next section, we find symmetry
transformations of a certain class of Markovian open-system evolutions which
are not symmetries of the corresponding no-jump trajectories. These transformations
are shifts of the Lindblad operators, i.e., of the form $L_m \rightarrow L_m -
f_m (t) \hat{1}$. Here, $f_m (t)$ are arbitrary complex-valued functions and are
hidden parameters in the sense that they do not affect this class of Markovian evolution
models. In Sec. \ref{sec:gp}, this result is illustrated by an explicit calculation
of the no-jump geometric phase for a dephasing qubit (spin$-\frac{1}{2}$) being
exposed to a static magnetic field. The geometric phase for stochastic unravelings
in the form of the linearized QSD equation is analyzed in Sec. \ref{sec:qsd}.
The paper ends with the conclusions.

\section{Shift symmetries of open-system evolutions}
\label{sec:shift}
We consider Markovian evolution of open quantum systems governed by
the Lindblad equation ($\hbar = 1$ from now on) \cite{lindblad76}
\begin{eqnarray}
\dot{\rho} (t) & = &
-i[H(t),\rho (t)] + \left. \lambda \sum_{m} \right( L_{m} \rho (t) L_{m}^{\dagger}
\nonumber \\
 & & \left. -
\frac{1}{2} L_{m}^{\dagger} L_{m} \rho (t) - \frac{1}{2} \rho (t) L_{m}^{\dagger}
L_{m}  \right)
\nonumber \\
 & = & -i[H(t),\rho (t)] + \lambda \mathcal{L} (\rho (t)) .
\label{eq:lindblad}
\end{eqnarray}
Here, $L_m$ are dimensionless Lindblad operators that model the influence of the
environment on the system evolution. For simplicity, we shall assume that $L_m$
are time-independent. The parameter $\lambda \geq 0$ controls the strength of
the open-system effect, such that $\lambda =0$ corresponds to unitary closed
system evolution.

The Lindblad equation obeys certain symmetries; an apparent one is the
independence of choice of zero point energy corresponding to the transformation
$H(t) \rightarrow H(t) - h(t) \hat{1}$, where $h(t)$ is real valued and $\hat{1}$
is the identity operator. Another general type of symmetry corresponds to the
transformation $L_m \rightarrow \sum_n V_{mn} L_n$, where $V_{mn}$ is an
arbitrary unitary matrix \cite{buric09}. One may check that this transformation
leaves the Lindblad equation unchanged and thus will not affect the state
$\rho (t)$ of the system.

The quantum jump unraveling is defined by dividing the evolution given
by Eq. (\ref{eq:lindblad}) into small time steps $\Delta t$. In the
$\Delta t \rightarrow 0$ limit, this procedure leads to quantum trajectories
in state space consisting of smooth deterministic parts interrupted
by random jumps, generated by jump operators proportional to $L_m$.
For a pure initial state $\psi_0$, these trajectories reside in projective
Hilbert space $\mathcal{P} (\mathcal{H})$ formed by rays of the system's
Hilbert space $\mathcal{H}$. These rays are equivalence classes consisting
of vectors that differ by multiplication of nonzero complex numbers. As shown in
Ref. \cite{carollo03}, one may associate a pure state geometric phase to
each such trajectory. Here, we focus on the geometric phase of no-jump
trajectories. Such a trajectory is the projection onto
$\mathcal{P} (\mathcal{H})$ of the continuous Hilbert space path
\begin{eqnarray}
[0,T] \ni t \rightarrow \ket{\psi(t)} =
{\bf T} e^{-i\int_0^t \widetilde{H}(t') dt'} \ket{\psi_0}
\end{eqnarray}
with ${\bf T}$ time ordering and
\begin{eqnarray}
\widetilde{H}(t) = H(t)-\frac{i}{2} \lambda \sum_m L_m^{\dagger}L_m
\label{eq:nojumpham}
\end{eqnarray}
a non-Hermitian effective no-jump Hamiltonian. The corresponding no-jump
geometric phase acquired on the time interval $[0,T]$ reads \cite{carollo03}
\begin{eqnarray}
\gamma_{nj} & = & \arg \langle \psi (0) \ket{\psi (T)} + \int_{0}^{T}
\frac{\bra{\psi(t)} H(t) \ket{\psi(t)}}{\langle \psi(t) \ket{\psi(t)}}dt .
\end{eqnarray}
Note that $\gamma_{nj}$ is a property of a path in $\mathcal{P} (\mathcal{H})$
as it is invariant under the transformation $\ket{\psi (t)} \rightarrow c(t)
\ket{\psi (t)}$ together with $H(t) \rightarrow H(t) + i\frac{d}{dt}
\ln \frac{c(t)}{|c(t)|}$, where $c(t)$ is a nonzero complex number for all $t\in [0,T]$.

It is straightforward to check that the no-jump
path $\ket{\psi(t)}$ and the corresponding geometric phase $\gamma_{nj}$
are unaffected by the above-mentioned unitary rotation $L_m \rightarrow
\sum_n V_{mn} L_n$. But there may be other symmetries that apply only
to certain classes of open systems. We focus on the symmetry related
to the shifts $L_m \rightarrow L_m - f_m(t)\hat{1}$ where $f_m (t)$
is in general complex-valued. Such shifts induce the transformations
\begin{eqnarray}
H(t) & \rightarrow & K(t) = H(t) - \frac{i}{2} \lambda \sum_m
\left( f_m^{\ast} (t) L_m - f_m (t) L_m^{\dagger} \right) ,
\nonumber \\
\mathcal{L} & \rightarrow & \mathcal{L} .
\label{eq:hamtransf}
\end{eqnarray}
Thus, they result solely in an extra term in the Hamiltonian part of
Eq. (\ref{eq:lindblad}). This implies that the Lindblad evolution is
unchanged under the shifts of $L_m$ if all $f_m^{\ast} (t) L_m$ are
Hermitian. In such a case, $f_m (t)$ are said to be hidden parameters
of the full open-system evolution. On the other hand, the no-jump
Hamiltonian transforms as
\begin{eqnarray}
\widetilde{H} (t) \rightarrow \widetilde{K}(t) & = & \widetilde{H} (t) +
\frac{i}{2} \lambda \sum_m \Big( f_m (t) L_m^{\dagger}
\nonumber \\
 & & + f_m^{\ast} (t) L_m -
|f_m (t)|^2 \hat{1} \Big)
\label{eq:nonhamtransf}
\end{eqnarray}
with $\widetilde{H} (t)$ the no-jump Hamiltonian in Eq. (\ref{eq:nojumpham}).
The transformed no-jump Hamiltonian $\widetilde{K} (t)$ may be nontrivially
different from $\widetilde{H} (t)$ even for Hermitian $f_m^{\ast} (t) L_m$.
Thus, for shifts $L_m \rightarrow L_m - f_m(t)\hat{1}$ such that
$f_m^{\ast} (t) L_m$ are Hermitian, the Lindblad evolution is unchanged
but the deterministic no-jump evolution may undergo a nontrivial change
originating from the anti-Hermitian contributions $\frac{i}{2}\lambda
\left( f_m(t) L_m^{\dagger} + f_m^{\ast} (t) L_m \right)$ to $\widetilde{K}(t)$.
In this case, the no-jump trajectories may depend on the parameters $f_m$,
which are hidden in the full open-system evolution. We note that this result
applies to any smooth portion of a quantum trajectory, i.e., trajectories
that contain one or several jumps share with the no-jump trajectories the
same kind of behavior under shifts of the Lindblad operators.

The $f_m$ dependence may be interpreted as an manifestation of a continuous
monitoring of the environment in the presence of a specific form of system-environment
interaction. To see this explicitly, let us consider a unitary representation model for the
system-environment evolution during the time interval $[t,t+\delta t]$, where
$\delta t$ is the finite time resolution for measuring projectively the environment
in some orthogonal basis $\{ \ket{0_e},\ket{m_e} \}$. We assume that $\delta t$
and $\lambda$ are much smaller than the typical energy shift associated with
$H(t)$. Under this assumption, the change in the system-environment state can
be described by a unitary map $U(t,t+\delta t; \{ f_m (t) \})$ with the effect
\begin{eqnarray}
\ket{0_e} \ket{\psi (t)} & \rightarrow & U(t,t+\delta t; \{ f_m (t) \})
\ket{0_e} \ket{\psi (t)}
\nonumber \\
 & = & \ket{0_e} \left( \hat{1} - i\widetilde{K} (t)
\delta t \right) \ket{\psi (t)}
\nonumber \\
 & & + \sqrt{\lambda \delta t} \sum_m \ket{m_e}
\left( L_m - f_m (t) \hat{1} \right) \ket{\psi (t)} .
\nonumber \\
\end{eqnarray}
Here, we have assumed that the environment is prepared in the pure state
$\ket{0_e}$ and we have taken
\begin{eqnarray}
\bra{0_e} U(t,t+\delta t; \{ f_m (t) \}) \ket{0_e} & = & \hat{1} - i\widetilde{K} (t)
\delta t ,
\nonumber \\
\bra{m_e} U(t,t+\delta t; \{ f_m (t) \}) \ket{0_e} & = & \sqrt{\lambda \delta t}
\left( L_m \right.
\nonumber \\
 & & \left. - f_m (t) \hat{1} \right)
\label{eq:U}
\end{eqnarray}
to the first order in $\delta t$ and $\sqrt{\lambda \delta t}$. Thus, the shifts
$L_m \rightarrow L_m - f_m (t) \hat{1}$ would correspond to engineering the system-environment
interaction so that Eq. (\ref{eq:U}) is satisfied. Evidently, the jump operators
are $\sqrt{\lambda \delta t} \left( L_m - f_m(t) \hat{1} \right)$.
The no-jump trajectory $[0,T] \ni t \rightarrow \ket{\psi (t)} =
{\bf T} e^{-i\int_0^t \widetilde{K} (t')dt'} \ket{\psi_0}$ is
realized with probability $\langle \psi (T) \ket{\psi (T)}$
by verifying that no change has occurred in the environment, and repeating
up to time $T$.

The operators $F_0 (t) = \hat{1} - i\widetilde{K} (t) \delta t$ and $F_m (t) =
\sqrt{\lambda \delta t} \left[ L_m - f_m (t) \hat{1} \right]$ in Eq. (\ref{eq:U})
constitute a set of Kraus operators that represent a completely positive map of
system states from $t$ to $t + \delta t$. Provided all $f_m^{\ast} (t) L_m$ are
Hermitian, there is a unitary matrix $\boldsymbol{W}(t)$ that relates this Kraus
representation with the original one consisting of $E_0 (t) = \hat{1} -
i\widetilde{H} (t) \delta t$ and $E_m (t) = \sqrt{\lambda \delta t} L_m$.
Explicitly, we may write $F_{\mu} (t) = \sum_{\nu} \boldsymbol{W}_{\mu\nu} (t)
E_{\nu} (t)$, $\mu,\nu = 0,1,\ldots$, with
\begin{widetext}
\begin{equation}
\boldsymbol{W} (t) =
\begin{pmatrix}
1-\frac{1}{2} \lambda \delta t \sum_m \left| f_m (t)\right|^2 &
\sqrt{\lambda \delta t} f_1^{\ast } (t) &
\sqrt{\lambda \delta t} f_2^{\ast } (t) & \sqrt{\lambda \delta t} f_3^{\ast } (t) & \ldots \\
- \sqrt{\lambda \delta t} f_1 (t) & 1 & 0 & 0 & \ldots \\
- \sqrt{\lambda \delta t} f_2 (t) & 0 & 1 & 0 & \ldots \\
- \sqrt{\lambda \delta t} f_3 (t) & 0 & 0 & 1 & \ldots \\
\vdots & \vdots & \vdots & \vdots & \ddots
\end{pmatrix} ,
\end{equation}
\end{widetext}
which can be checked to be unitary to lowest order in $\delta t$ and $\sqrt{\lambda \delta t}$.
The existence of such a unitary matrix demonstrates \cite{nielsen00} that the two maps
are physically identical if all $f_m^{\ast} (t) L_m$ are Hermitian; a result that is
consistent with the invariance of the Lindblad equation under the
shifts $L_m \rightarrow L_m - f_m (t) \hat{1}$ for this type of open-system evolution.

\begin{figure}[htb]
\includegraphics[width = 8 cm]{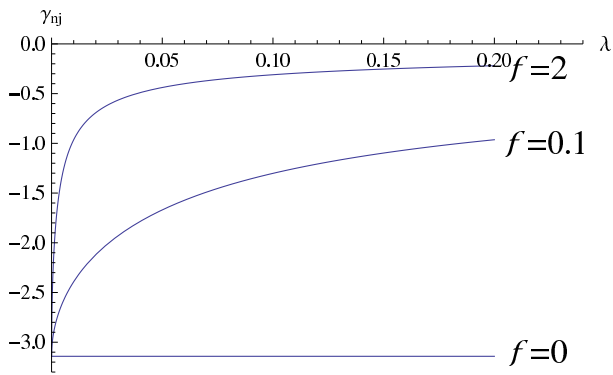}
\caption{\label{fig:1} (Color online) The no-jump phase $\gamma_{nj}$ as a function of
the open-system strength $\lambda$ for the shift values $f=0,0.2,2$. The
initial state is taken to be pure on the equator ($\theta_0 = \frac{\pi}{2}$)
of the Bloch sphere. The horizontal curve for $f=0$ demonstrates the
$\lambda$ independence found in Refs. \cite{carollo03,fuentes05}. For
$f\neq 0$, $\gamma_{nj}$ depends strongly on $\lambda$.}
\end{figure}

\section{Geometric phase}
\label{sec:gp}
We next show that the previous result may have consequences for the geometric
phase of a no-jump trajectory. We find an explicit physical example where
a shift parameter $f$ is imprinted in the no-jump geometric phase, although
the corresponding open-system evolution is $f$-independent.

Consider a qubit (spin$-\frac{1}{2}$) prepared in the pure state $\ket{\psi_0} =
\cos \left( \frac{1}{2} \theta_0 \right) \ket{0} + \sin \left( \frac{1}{2} \theta_0
\right) \ket{1}$, exposed to a static magnetic field in the $z$
direction and to dephasing of strength $\lambda$. This may be modeled
by a Hamiltonian $H = \frac{\omega}{2}\sigma_z$ and a single Lindblad operator
$L = \sigma_z$, with $\omega$ the precession frequency and $\sigma_z =
\ket{0} \bra{0} - \ket{1} \bra{1}$ the $z$ component of the standard Pauli
operators. By using Eqs. (\ref{eq:hamtransf}) and (\ref{eq:nonhamtransf}), we
find the transformations $H \rightarrow K = H - \lambda \im (f) \sigma_z$,
$\mathcal{L} \rightarrow \mathcal{L}$, and
$\widetilde{H} \rightarrow \widetilde{K} = \widetilde{H} + i\lambda \re (f)
\sigma_z - \frac{i}{2} \lambda |f|^2 \hat{1}$, under the shift
$L \rightarrow L - f\hat{1}$, where $f$ is assumed to be time-independent
for simplicity. Thus, the Lindblad evolution is unaffected by the shift if
$f$ is real-valued. On the other hand, the geometric phase of the no-jump
trajectory
\begin{eqnarray}
[0,T] \ni t \rightarrow \ket{\psi (t)} = e^{-i \widetilde{K} t} \ket{\psi_0}
\label{eq:dynamics}
\end{eqnarray}
for a quasi-cyclic path over the time interval $[0,2\pi/\omega]$ with
real-valued $f$, takes the form
\begin{eqnarray}
\gamma_{nj} & = & -\pi + \frac{\omega}{4f \lambda} \ln \left(
e^{\frac{4\pi}{\omega} f \lambda} \cos^2\frac{\theta_0}{2} \right.
\nonumber \\
 & & \left. + e^{-\frac{4\pi}{\omega} f \lambda}
\sin^2\frac{\theta_0}{2} \right) ,
\label{eq:njgp}
\end{eqnarray}
which is explicitly $f$-dependent.

The geometrical reason for this $f$-dependence can be seen by looking at the
Bloch sphere polar angle $\theta$, which becomes time-dependent if $f\neq 0$.
Explicitly, by evaluating the right-hand side of Eq. (\ref{eq:dynamics}) we
obtain $\tan [\theta (t)/2] = e^{-2f\lambda t} \tan (\theta_0/2)$ and azimuthal
angle $\varphi (t) = \varphi_0 + \omega t$, which correspond to a spiralling
motion toward the north (south) pole of the Bloch sphere for $f>0$ ($f<0$)
and all $\theta_0 \neq \pi$ ($\theta_0 \neq 0$). Furthermore, one may check
that $\gamma_{nj}$ in Eq. (\ref{eq:njgp}) converges to the expected
$-\pi(1-\cos\theta_0)$ (minus half the solid angle enclosed on the Bloch sphere)
in the $f\lambda \rightarrow 0$ limit. The nontrivial $f$ dependence is
illustrated in Fig. \ref{fig:1}. The resilience to dephasing of the geometric
phase of the no-jump trajectory found in Refs. \cite{carollo03,fuentes05}
corresponds to the case $f=0$. However, as our calculation shows, any nonzero
$f$ would predict $\gamma_{nj}$ to be $\lambda$-dependent and thus be affected by
this kind of open-system effect. For small $f\lambda$, this dependence is
linear, which may be seen by expanding the geometric phase around the closed
system expression, leading to the lowest order correction
$2\pi^2 \frac{f\lambda}{\omega} \sin^2 \theta_0$.

We may show that the no-jump evolution of dephasing for $f \neq 0$ is equivalent
to decay of the precessing qubit. Consider the evolution generated by the Lindblad
operator $L_- = \sigma_x - i\sigma_y$, which corresponds to decay toward the south
pole of the Bloch sphere with some strength $\lambda'$, say. The no-jump curve is
determined by the effective no-jump Hamiltonian $\widetilde{H}' = \frac{\omega}{2}
\sigma_z - i\lambda' \sigma_z -i\lambda' \hat{1}$, where we have assumed the Hamiltonian
$H = \frac{\omega}{2} \sigma_z$. If $\lambda' = -f\lambda$, then the no-jump
Hamiltonian $\widetilde{H}'$ generates the same curve in $P(\mathcal{H})$ as
$\widetilde{H}$ in the dephasing model (the corresponding Hilbert space curves
differ only by multiplication of a nonzero complex number).

It is instructive to compare the preceding dephasing example with the mixed state
geometric phase $\gamma [\mathcal{P}]$ proposed in Ref. \cite{tong04} for
real-valued $f$. Since $\gamma [\mathcal{P}]$ is based directly on the
kinematics $\rho (t)$ of open-system evolution, it follows immediately
that $\gamma [\mathcal{P}]$ is $f$-independent and therefore experimentally
testable \cite{cucchietti10}. It was furthermore found that $\gamma [\mathcal{P}]$
for precession around the $z$ axis is resilient to dephasing only if the
initial state lies on the equator of the Bloch sphere. This particular form
of resilience has been reported in a recent experiment \cite{filipp09} with
polarized ultracold neutrons exposed to dephasing noise.

For one-qubit systems, $f_m^{\ast} (t) L_m$ can be chosen to be Hermitian for
all open-system models that include Lindblad operators that are linear combinations
of the Pauli operators with real coefficients, such as dephasing and depolarization.
On the other hand, it should be stressed that if no $f_m (t)$ exist such that
$f_m^{\ast} (t) L_m$ becomes Hermitian, then the shift in fact corresponds to a
new Hamiltonian that may cause a different evolution $\rho (t)$. One such
example is spontaneous decay of a qubit, for which one cannot find a nonzero $f$
such that $f^{\ast}L_{-}$ becomes Hermitian. Indeed, the shift $L_{-} \rightarrow L_{-} -
f\hat{1}$ induces the Zeeman term $\lambda \left[ \im (f) \sigma_x + \re (f)
\sigma_y \right]$ in the Hamiltonian, in the case of spontaneous decay. Another
example is a spin$-\frac{1}{2}$ system interacting with a quantized light field
and subjected to a linear loss of photons. This loss may be modeled by the non-Hermitian
photon annihilation operator $L=a$. In this case, the shift $L \rightarrow L - f\hat{1}$
yields the extra term $\sqrt{2} \left[ -\im (f) x + \re (f) p \right]$ in the
Hamiltonian, where $x=(a+a^{\dagger}) / \sqrt{2}$ and $p=(a-a^{\dagger}) /
(i\sqrt{2})$ are the `position' and `momentum' operators, respectively, of
the quantized light field. It is reasonable to expect that any robustness in
the geometric phase found in these two latter types of systems would signal an
error resistance also at the level of the full Lindblad evolution. Applications
of the quantum jump unraveling to these model systems have indeed shown a
nontrivial error resilience \cite{cen04,carollo04b}, results that support
the usefulness of the geometric phase for robust quantum computation.

\section{Stochastic unravelings}
\label{sec:qsd}
Stochastic unravelings in the form of quantum state diffusion (QSD) \cite{gisin92}
consist of continuous, Brownian-like quantum trajectories whose average coincides
with the full Lindblad evolution. The geometric phase of such trajectories arising
from nonlinear \cite{gisin92} and linear \cite{goetsch94} versions of the
QSD equation has been considered in Refs. \cite{bassi06} and \cite{buric09},
respectively. Reference \cite{bassi06} considered the phase transformation
$L_m \rightarrow e^{i\chi_m} L_m$ and found a nontrivial $\chi_m$ dependence
in the geometric phase for the nonlinear QSD evolution. In Ref. \cite{buric09},
it was demonstrated that the averaged geometric phase $\alpha_g$ associated
with the linearized evolution is invariant under unitary rotations
$L_m \rightarrow \sum_n V_{mn} L_n$, provided the system starts in a pure
state. Here, we examine the behavior of this geometric phase under the shifts
$L_m \rightarrow L_m - f_m (t) \hat{1}$ and show that $\alpha_g$ may depend on
$f_m$, also when $f_m$ is hidden in the full open-system evolution.

The linearized QSD equation reads
\begin{eqnarray}
\ket{d\phi} & = & \Big[ -i H(t)dt - \frac{1}{2} \lambda \sum_m L_m^{\dagger} L_m
\nonumber \\
& & + \sqrt{\lambda} \sum_m L_m dw_m \Big]
\ket{\phi},
\end{eqnarray}
where $w_m$ are complex Wiener processes with respect to a probability measure
$\mathbb{Q}$. There is a mean $\mathbb{E}_{\mathbb{Q}}$ over $\mathbb{Q}$ such
that $\mathbb{E}_{\mathbb{Q}} [dw_m] =\mathbb{E}_{\mathbb{Q}} [dw_mdw_{m'}] = 0$;
$\mathbb{E}_{\mathbb{Q}} [dw_m dw_{m'}^{\ast}] = \delta_{mm'} dt$. This guarantees
the properly renormalized average of any measurable quantity to coincide with the
expectation value with respect to $\rho (t)$. Following Ref. \cite{buric09}, the
averaged geometric phase $\alpha_g$ with respect to the probability measure
$\mathbb{Q}$ is taken to be
\begin{eqnarray}
\alpha_g = \arg \mathbb{E}_{\mathbb{Q}} [\langle \phi_0 \ket{\phi (T)}] +
\int_0^T \Tr \left[ \rho (t) H(t) \right] dt .
\end{eqnarray}
The second term on the right-hand side of this expression depends only on the
full state $\rho (t)$ and would therefore be unaffected under all symmetry
transformations of the Lindblad equation. To show the noninvariance of
$\alpha_g$ under shifts of the Lindblad operators, it is therefore sufficient
to show that the first term may be $f_m$ dependent. We demonstrate this by
an example, again the dephasing qubit model with real-valued and
time-independent shift parameter $f$ and Hamiltonian $H=\frac{\omega}{2}\sigma_z$.
For initial state $\ket{\phi_0} = \cos \left( \frac{1}{2} \theta_0 \right) \ket{0} +
\sin \left( \frac{1}{2} \theta_0 \right) \ket{1}$, we obtain
\begin{widetext}
\begin{eqnarray}
\arg \mathbb{E}_{\mathbb{Q}} [\langle \phi_0 \ket{\phi (T)}] = \arg \bra{\phi_0}
\exp \left[ -i \left( \frac{1}{2} \omega + if\lambda \right) T \sigma_z \right] \ket{\phi_0} =
-\arctan \left[ \frac{\tanh (f\lambda T) + \cos \theta_0}{1+\tanh (f\lambda T)
\cos \theta_0} \tan \left( \frac{\omega T}{2} \right) \right] .
\end{eqnarray}
\end{widetext}
Thus, $\arg \mathbb{E}_{\mathbb{Q}} [\langle \phi_0 \ket{\phi (T)}]$
is $f$ dependent if $\omega T \neq n\pi$, $n$ integer, and $\cos \theta_0
\neq \pm 1$. Thus, it follows that the averaged geometric phase $\alpha_g$
associated with the linearized QSD evolution may depend on the hidden
parameter $f$.

\section{Conclusions}
\label{sec:conclusions}
We have demonstrated the existence of Markovian open-system evolutions for
which the associated no-jump quantum trajectories may depend on parameters
that are undetermined by the full open-system evolution. We
have found conditions for this situation to occur and have identified
the origin of this dependence in terms of continuous monitoring of
the system's environment. Furthermore, we have explicitly demonstrated
how such a hidden parameter can be unveiled by the geometric phase of an
individual quantum trajectory for a dephasing qubit. The realization
of the geometric phase for single quantum trajectories requires explicit
engineering of the system-environment interaction; a feature that is shared
by the mixed state geometric phases for completely positive maps proposed
in Ref. \cite{ericsson03}. Finally, we have demonstrated that the averaged
geometric phase introduced in Ref. \cite{buric09} of the linearized QSD
model shows a similar dependence on hidden parameters. Thus, it remains
open whether a well-defined open-system geometric phase based upon quantum
trajectories exists.
\vskip 0.3 cm
We would like to thank Marcelo Fran\c{c}a Santos and Vlatko Vedral for
useful discussions, and Dianmin Tong for comments on the manuscript.
E.S. acknowledges support from the National Research Foundation and the
Ministry of Education (Singapore).

\end{document}